\documentstyle[aps,epsfig]{revtex}
\begin{document}
\twocolumn
\title{
Low energy measurement of the $^7$Be(p,$\gamma$)$^8$B cross section
}
\author{F. Hammache$^{1,*)}$, G. Bogaert$^{1,**)}$, P. Aguer$^{2)}$, C.
Angulo$^{3)}$, S. Barhoumi$^{5)}$,
L. Brillard$^{4)}$, J.F.~Chemin$^{2)}$, G. Claverie$^{2)}$, A. Coc$^{1)}$, M.
Hussonnois$^{4)}$, M. Jacotin$^{1)}$, J. Kiener$^{1)}$, A.~Lefebvre$^{1)}$,
C.~Le~Naour$^{4)}$, S. Ouichaoui$^{5)}$, J.N. Scheurer$^{2)}$, 
V. Tatischeff$^{1)}$,
J.P. Thibaud$^{1)}$, E.~Virassamyna\"{\i}ken$^{2)}$}
\address{
$^{1)}$ CSNSM, IN2P3-CNRS, 91405 Orsay et Universit\'e de Paris-Sud, France\\
$^{2)}$ CENBG, IN2P3-CNRS, et Universit\'{e} de Bordeaux I, 33175 Gradignan,
France\\
$^{3)}$ Centre de Recherches du Cyclotron, 
UCL, Chemin du Cyclotron 2, 1348 Louvain la Neuve, Belgium\\
$^{4)}$ IPN,IN2P3-CNRS et Universit\'{e} de Paris-Sud, 91406 Orsay, France\\
$^{5)}$ Institut de Physique, USTHB, B.P. 32, El-Alia, Bab Ezzouar, Algiers,
Algeria}

\maketitle
\begin{abstract}
We have measured the cross section of the $^{7}$Be(p,$\gamma$)$^{8}$B
reaction for E$_{cm}$ = 185.8 keV,
134.7 keV and 111.7 keV using a radioactive $^7$Be target (132 mCi).
Single and coincidence spectra of 
$\beta^{+}$ and $\alpha$ particles from $^{8}$B and $^{8}$Be$^*$ decay, 
respectively,
were measured
using a large acceptance spectrometer.
The zero energy S factor inferred from these data is 18.5 $\pm$ 2.4 eV b and
a weighted mean value of 18.8$\pm$1.7 eV b (theoretical uncertainty included) is
deduced when combining this value with our previous results at higher energies. 
\end{abstract}

\bigskip
PACS numbers : 25.40.LW, 27.20.+n, 26.65.+t\\


Recent experimental results on solar $\nu_e$ and 
atmospheric  $\nu_\mu$ neutrinos
support neutrino oscillation scenarios, in which the
oscillation probability depends on the neutrino mixing angles and squared
mass differences. For $\nu_e$-$\nu_x$ oscillation, the determination of these
fundamental quantities needs accurate solar modeling and nuclear cross sections
for the reactions operating
in the solar core \cite{BAH99,BRU98,MOR99}.
In this respect,
the most important nuclear physics parameter is
the S factor of the $^7$Be(p,$\gamma$)$^8$B reaction
which gives rise to the crucial $^8$B neutrinos \cite{SCH99}.
In a previous work \cite{HAM98}, we have measured the
$^7$Be(p,$\gamma$)$^8$B reaction  cross section for E$_{c.m.}$ = 0.35 - 1.4 MeV using
radioactive $^7$Be targets. 
In this Letter, we report on new direct measurements of this cross
section at center of mass energies below 200 keV, where extrapolation
to solar energies ($E_{cm} = $ 18 keV) is
expected to be almost free of theoretical uncertainties \cite{ADE98}, which is not the case for measurements at higher energies. 

The electrostatic accelerator PAPAP
at Orsay  supplied intense proton beams of well calibrated energies 
\cite{BOG94}.
We used a highly radioactive $^7$Be target (131.7 mCi)
prepared as in ref.
\cite{{HAM99},{HUS00}} additionally containing  approximately 3.10$^{16}$ atoms of
$^9$Be.
$\beta^+$ and $\alpha$ particles from
$^8$B($\beta^+$)$^8$Be$^*$(2$\alpha$) decays were detected at forward 
and backward angles respectively. 
We used the solenoidal superconducting
magnet SOLENO \cite{SOL}(3.2 T at the center, 1.22 m long, 32 cm of internal 
diameter) in order to detect 
 both $\alpha$  and 
$\beta^+$ particles with high efficiency (11.5 \% and 25\% respectively) due to
the focusing power of  the field.
Both singles and coincidence
events between $\alpha$ and $\beta^+$ particles were recorded, the latter
providing spectra free of background  events even  at low
bombarding energies.

Off beam detection ($^8$B period = 0.77 s) was necessary to avoid any contamination in both $\alpha$
and $\beta^+$ coincidence spectra arising
from the channel $^7$Li(p,$\gamma_1$)$^8$Be$^*$(2$\alpha$). In this reaction the 
$\alpha$ decay of $^8$Be$^*$ is the same as in
$^8$B($\beta^+$)$^8$Be$^*$(2$\alpha$) and
the $\beta^+$  detector could not efficiently discriminate
$\beta^+$ particles of interest from electron pairs produced in the target
assembly by  $\gamma_1$ rays (E$_{\gamma_1}$ =14.8 MeV).

For the delayed detection purpose, the beam passed through an electrostatic 
deflector
which was alternately
switched on and off for  time periods of 1.5 s.
The prompt events from $^9$Be(p,$\alpha$)$^6$Li and
$^9$Be(p,d)$^8$Be were also recorded and used later for energy calibration
of the $\alpha$ detectors and 
normalization purpose (see below).

The target was located near the solenoid center, perpendicular to the symmetry axis of the field. 
The target backing of 0.1 mm ultra pure platinum and its holder  
allowed both
efficient water cooling and large transmission  for $\beta^+$ particles. 
The beam spot at the target position was made visible
by using an alumina ($Al_2O_3$)target 
which could be moved to the
exact position of the target. 
An optical system provided  a magnified image of the beam spot 
to  measure its position and size on the target. They were 
systematically determined (size about 10 mm$^2$) before and after each run.

In order to prevent carbon buildup on the target, a copper plate
cooled with liquid nitrogen was placed in the target vicinity.
In addition, 
cryogenic, turbomolecular and ionic pumps were used to keep 
the vacuum  below 5 $\times$ 10$^{-7}$ mbar during the whole experiment.

Typical beam currents of 10-40 $\mu$A on the target were used. To suppress
secondary electron escape, the target was biased at a positive 300 V. 
As a consistency check, currents
were also  measured in the insulated  chamber where the beam was periodically 
deflected, at the SOLENO entrance collimator and 
at the $\alpha$ detector tube (see below) where negligible 
currents were observed. All the currents were integrated, digitized, 
recorded cycle-by-cycle and analyzed off-line. Very good  agreement was found
between integrated
charges for beam on target and beam off target, and the overall uncertainty on the 
target integrated charge was 2 \%. 

$\beta^+$ particles (E$_{max}$ = 14 MeV) were detected in a set of
6 successive cylindrical plastic scintillators (diameter 20 mm and thickness 
3, 3, 8, 8, 8 and 10 mm) 
centered on the field
axis and 22 cm
away from the target. The number and thickness
of the plastic scintillators were optimized from GEANT simulations
\cite{GEANT}
to discriminate MeV $\beta^+$ particles 
from the huge number of $\gamma$ rays and photoelectrons produced in the target
backing by the
$E_{\gamma}$=478 keV radiation from $^7$Be decay while still measuring 
$\beta^+$  energies with a reasonable precision. This discrimination was
effective when at least two $\beta$ detectors were required to fire.

$\alpha$ particles (from $^8$Be$^*$ decay) with energies below  3.5 MeV 
and emission angles between approximately 95  and
150 degrees with respect to the beam direction were deflected towards the
solenoid axis and
detected in an array of 6 $\times$ 4  Si detectors (22 mm $\times$ 45 mm 
$\times$ 0.1 mm). 
The detectors were 
mounted in  cylindrical geometry (internal diameter of 4 cm) aligned on the solenoid axis.
Depending on energy and emission angle, $\alpha$  particles were detected 
at distances  ranging from 18 to 36.5 cm from the target.
Backscattered protons were deflected along different paths and were not able to reach the 
$\alpha$ detectors. 

The data were recorded event by event. They included the measurements of $\alpha$ and $\beta^+$ 
energies, 
number and identification of detectors fired and the time of flight difference between $\alpha$ and 
$\beta^+$ particles 
for coincidence events. 
In the data analysis, events where more than one $\alpha$ detector or less than two $\beta^+$ 
detectors fired were rejected. 
A pulse generator was used for dead time measurements.

\begin{figure}
\vspace{-1.0 cm}
\begin{center}

\mbox{\epsfig{file=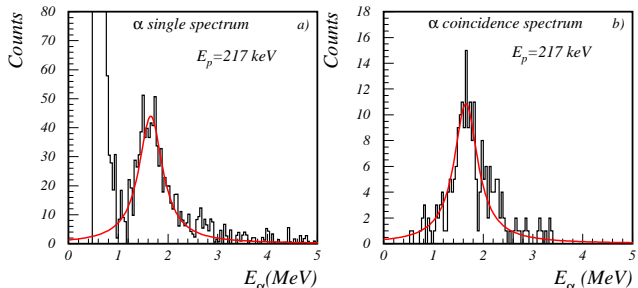,width=9cm}}
\caption{a). Energy spectrum
of singles delayed $\alpha$
particles. 
b). Energy spectrum of delayed $\alpha$
particles detected in coincidence
with delayed $\beta$ particles. }
\end{center}

\end{figure}

Cross section measurements were performed at three proton energies,
217 keV, 160 keV and 130 keV.
The absolute value
of the cross section at the proton energy of 217 keV was obtained counting
singles delayed $\alpha$ particles in the range from 1 MeV to 3.36 MeV.
Figure 1.a. shows the corresponding 
spectrum,  obtained by adding together the individual $\alpha$ spectra
measured in the silicon detectors. 

The calibration
of the silicon detectors was performed using the three well defined peaks observed
in the singles prompt spectra from $^9$Be(p,d)$^8$Be and
$^9$Be(p,$\alpha$)$^6$Li reactions.
Data shown in figure 1.a were obtained after subtraction of the background
contribution  which was determined in a series of measurements of several days
with the beam off and the active target in  place. A check was made to
ensure that a null extra background (within statistical uncertainties) was 
introduced when the beam hit a pure platinum target.
Background subtraction was performed by simply normalizing the
corresponding spectrum to counting time. 
For comparison, the coincidence delayed
$\alpha$ particle spectrum measured in the same runs is shown in figure 1.b.
It can be
seen in figure 1 that the low energy ($<$ 1 MeV) component  in the
singles spectrum, corresponding to pileup events due to photoelectrons
created by the 478 keV $\gamma$ rays, has completely vanished in the
coincidence spectrum. 
The solid curve in figure 1.b is obtained from a
least squares fit to this  background free
coincidence spectrum. We see in  figure 1.a that 
the same curve also
 provides  a reasonable fit to singles data 
(after normalization to counting), as expected 
from an unbiased background subtraction process.

The $\alpha$ detection efficiency in the 1.00-3.36 MeV energy range was
determined with the same detection
setup  from an analysis of the reaction
$^7$Li(p,$\gamma_1$)$^8$Be$^*$ performed with an enriched $^7$Li target at E$_p$ = 160 keV.
As explained above, the 14.8 MeV $\gamma_1$ rays create e$^+$-e$^-$ pairs in the target
backing which were detected in the plastic scintillator counter,
while  $^8$Be$^*$  decay into two $\alpha$ particles detected in 
the silicon array (the same $\alpha$'s as in the channel
$^8$B($\beta^+$)$^8$Be$^*$). The $\alpha$ detection efficiency was deduced from
the number of counts in the plastic scintillator taken in singles and in
coincidence with the silicon detector. 
The singles
$\gamma_ 1$ ray yield was taken to be 0.680 $\pm$ 0.043 \cite{ZAH95} times the total 
counts arising in the plastic scintillator from both 
$\gamma_0$  and $\gamma_1$ channels.
GEANT simulations confirmed that pair detection efficiencies were
the same, within less than 1\%, for $\gamma_0$ and $\gamma_1$. The
angular correlation between e$^+$-e$^-$ pairs and $\alpha$ particles was
also calculated and found negligible for the s-wave proton capture  at
this low bombarding energy. 

A kinematics correction of 9.6\% was applied to the experimental
efficiency value
to take into account the $^8$Be$^*$
in flight-decay  in the
$^7$Li(p,$\gamma_1$)$^8$Be$^*$
reaction. 
Finally, 
the detection efficiency was $\epsilon_\alpha$ 
= 0.115 $\pm$ 0.008 in the 1.00 - 3.36 MeV energy range, in fair agreement 
with GEANT simulations (12 \%).

The $^{7}$Be total activity at the beginning and at the end of the run at 217 keV,
the target area (0.47 $\pm$ 0.02 cm$^2$) and the $^7$Be
activity profile were accurately determined with the same instruments and methods
used  in the experiment described in ref. \cite{HAM98}. After fitting the $^7$Be decay 
function to the measurements, we found an initial total activity of 
131.7 $\pm$ 2.4  mCi. No loss of activity due to beam impact was observed during the run.
The $^{7}$Be areal density over the beam spot was finally determined
run by run (9 \% uncertainty) by averaging the results of the 478 keV 
$\gamma$ ray scan over the beam spot dimensions and normalizing them to the 
total activity per unit surface area. 

A value of $\sigma$ = 16.7 $\pm$ 2.1 nb at the incident proton energy of 217 keV 
was deduced (see the formula 3 in ref. \cite{FIL83}).
This value takes into account a 1\%
correction due to the $^8$B backscattering on platinum atoms and escape out of the target \cite{WEI98}.
This correction was calculated using a TRIM \cite{TRI} simulation
 with target thickness and composition determined from   
consistent RBS, (d,p) and PIXE analysis measurements performed during 
the course of the experiment.
At the beginning of the experiment, the target was found to 
contain mainly carbon (9 $\mu$g/cm$^2$), oxygen 
(7.6 $\mu$g/cm$^2$) and less than 4 $\mu$g/cm$^2$ of calcium and 
other lighter elements, corresponding to a thickness of 
9.6 $\pm$ 1.0 keV for protons of 217 keV. 

This thickness leads to an effective energy of 212.4 keV (E$_{cm}$ = 185.8 keV)
and an S factor value of  17.2 $\pm$ 2.1 eV b.
The quoted uncertainty includes the 6.3\% uncertainty in the $\gamma_1$
branching ratio \cite{ZAH95} of $^7$Li(p,$\gamma$)$^8$Be$^*$.

The cross sections at E$_p$ = 130 keV and 160 keV were determined 
using $\alpha$-$\beta$ coincidence measurements only, 
because of the decrease of the signal over background ratio observed in singles 
spectra with lowering bombarding energies. The corresponding coincidence
$\alpha$ energy spectra are shown in figure 2 together with time difference
spectra between $\alpha$ and $\beta^+$ particles.
The 3 peaks in the time  spectra correspond to 3 classes of 
trajectories where $\alpha$ particles can spiral 1,2 or 3 times in the magnetic field before reaching the detectors.  The rare counts found between the peaks and at 
$\Delta ${\it t}  = 200 ns (i.e. null time of flight difference) in figure 2.b 
are background events (most
probably cosmic rays) eliminated in the energy spectra by gating on 
the three time peaks.

\begin{figure}
\vspace{-1.5 cm}
\begin{center}
\mbox{\epsfig{file=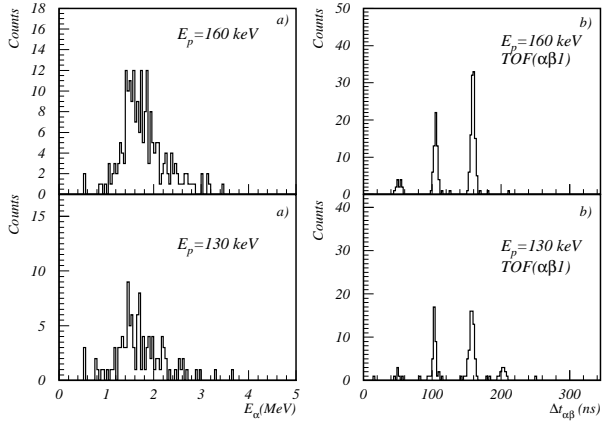,width=9cm}}
\caption{a). Energy spectra obtained at 160 keV and 130 keV
for delayed $\alpha$ particles detected in coincidence
with delayed $\beta$ particles. b).
Corresponding spectra for time difference between $\alpha$  and
$\beta^+$ particles
obtained using the first of the six  plastic scintillators.
A null time of flight
difference is arbitrarily at 200 ns due to delays in the electronics.}
\end{center}
\end{figure}

S factors at E$_p$ = 160 keV and E$_p$ = 130 keV, relative to the one measured 
at E$_p$ = 217 keV, were obtained by normalization to the $\alpha$ yield from
the reaction $^9$Be(p,$\alpha$)$^6$Li through the relation:

\begin{eqnarray}
\frac{S^{7}(E_{cm_{2,3}}^{7})}{S^{7}(E_{cm_1}^{7})}
=K \frac{R(E_{2,3})^{7,9}}
{R(E_{1})^{7,9}}
\frac{S^{9}(E_{cm_{2,3}}^{9})}
{S^{9}(E_{cm_1}^{9})}
 e^{-\lambda^{^7Be}(\Delta t_{1i})}
\end{eqnarray}

where the subscripts 1, 2 and 3 label  the runs at E$_p$=
217 keV, 160 keV and 130 keV, respectively, and the superscripts 7 and 9 label
the reactions $^7$Be(p,$\gamma$)$^8$B and $^9$Be(p,$\alpha$)$^6$Li, respectively.
$R(E_i)^{7,9}$ is the coincidence yield  normalized to the $\alpha$ yield
from the $^9$Be(p,$\alpha$)$^6$Li reaction, $S^{9}$ is the astrophysical S factor
of the $^9$Be(p,$\alpha$)$^6$Li reaction at the corresponding
c.m. effective energy. $K$ is a
constant accounting for the changes in dead times, in effective time parameters
(see parameter $\beta$ in formula 4 of ref. \cite{FIL83}) and in angular
distributions of alphas in $^9$Be(p,$\alpha$)$^6$Li with the bombarding
energies. These 3  corrections were found to be very small (less than a few
percent). The exponential term, in which $\Delta t_{1i}$ is the time difference
 between experiment i and 1, accounts for the decrease in the $^7$Be
target activity with time.

In calculating $R$, the $\alpha$ yield from $^9$Be(p,$\alpha$)$^6$Li
resulted from a least-squares analysis of the 
prompt singles spectra where very well defined $\alpha$ peaks show-up. 
S factors and 
angular distributions concerning the $^9$Be(p,$\alpha$)$^6$Li reaction 
were taken from the literature \cite{ZAH95}.

It must be stressed that the normalization to the $\alpha$ yield from
$^9$Be(p,$\alpha$)$^6$Li eliminates effects due to target non uniformity, beam
position variation and loss of activity of the target due to beam 
impact as long as the ratio of atomic densities of $^9$Be 
to  $^7$Be remains constant. This is expected since 
 $^9$Be was introduced in the $^7$Be solution before
electrodeposition of the final target. It was verified experimentally 
through a comparison of $^7$Be $\gamma$ ray scan with a $^9$Be scan using (d,p)
reaction analysis with a microbeam. A non negligible loss of activity was
actually observed after the runs at E$_p$= 160 keV and 130 keV,
because of a significant increase of sputtering effects with decreasing
energy.

For the calculations of the proton energy losses at 160 and 130 keV, we took
into account the loss of target material by the monitoring of the $^9$Be
content through the $^9$Be(p,$\alpha$)$^6$Li reaction. We estimated the
uncertainty on the effective energy to be 2.5 keV, which induces an
uncertainty of 2\% on the ratio 
$S^{9}(E_{cm_{2,3}}^{9})/ S^{9}(E_{cm_1}^{9}) $.

Taking into account  a 1\% loss of $^8$B nuclei due to the $^8$B backscattering
\cite{WEI98} 
we deduced the astrophysical S factor values of 
S(134.7 keV) = 19.5 $\pm$ 3.1 eV b  and S(111.7 keV) = 15.8 $\pm$ 2.7 eV b.
Final  uncertainties were calculated by quadratic summation of all individual
uncertainties related above.

Results for the astrophysical S factor are shown in figure 3. 
Extrapolation to zero energy using the calculations of ref. \cite{DES94} and 
the present low-energy data gives S(0) = 18.5 $\pm$ 2.4 eV b where the error bar is 
only experimental. A negligible dispersion of S(0)  
is found (0.2 eV b) when various calculated curves \cite{{DES94},{JOH92}}
 of S(E) are fitted to the same data. This reflects the agreement 
between models at low energy, since  the interaction in that case takes place at 
very large $^7$Be-p distances and is mainly 
governed by Coulomb physics. As a consequence the total uncertainty 
(experimental + theoretical) is finally $\pm$ 2.4 eV b.

 Using the calculation of ref. \cite{DES94}, our previous measurements \cite{HAM98} 
lead to S(0)=19.1 $\pm$ 1.2 eV b \cite{COR}, where the error bar is experimental.
 As the experiment was performed at higher energies,   
 sophisticated nuclear calculations are required in that case
 to describe the shape of S(E) which leads to a higher theoretical uncertainty on S(0). 
An S(0) dispersion of 
$\pm$ 2 eV b was found using the available models \cite{{DES94},{BAR95}} to fit the data. Adding 
this dispersion, considered as a reasonable estimate of the theoretical uncertainty, to the 
experimental error bar and combining quadratically the obtained result for S(0) with that of the 
present low-energy result, we finally obtain a weighted mean value of 
S(0)= 18.8 $\pm$ 1.7 eV b (taking a more  conservative value of $\pm$3 eV b for 
the theoretical uncertainty would lead to a very similar result of 18.7$ \pm$ 1.9 eV b.).

\begin{figure}
\vspace{-1.0 cm}
\begin{center}
\mbox{\epsfig{file=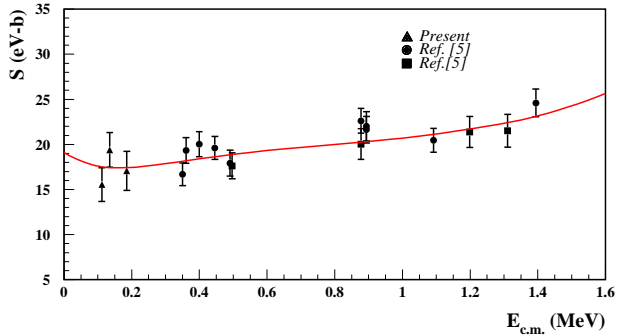,width=9cm}}
\caption{Measured S factors
from the  present work and from reference [5]
after backscattering
correction. [5]). Error bars represent random uncertainties.
The curve through the data, given for illustrative purposes, is a fit to the three sets of data, 
assuming independent errors and using  the calculation of ref. [16].}
\end{center}
\end{figure}

These results are in good agreement with some of the previous direct 
experiments \cite{FIL83,VAU70,HAS99}, (see \cite{CHE} for a comment on the recoil 
nuclei escape in refs. \cite{{FIL83},{VAU70}}). Concordant results but
with larger uncertainties have also been reported in recent studies
of the inverse process  \cite{{KIK98},{IWA99}} and of transfer reactions \cite{AZH99}. 

The present result, which includes a total uncertainty significantly lower 
 and better-founded than in  higher-energy measurements should help in
 clarifying the interpretation of solar neutrino experiments.

J.J. Correia, R. Daniel, D. Linget, N. Karkour are gratefully acknowledged for 
experimental support. Valuable discussions with M. R. Haxton are greatly acknowledged. 
This work was supported in part by Region Aquitaine.


\begin{thebibliography}{20}
\bibitem{COR}   S(0) = 19.1 $\pm$ 1.2 eV b is obtained when experimental 
values  are corrected for the $^8$B escape effect \cite{WEI98}. 
S(0) = 18.5 $\pm$ 1.0 eV b was deduced from uncorrected data in reference \cite{HAM98}
\bibitem{CHE} In these measurements, escape of recoil nuclei out of the
target was not corrected, and cannot be accurately calculated  since the
detailed compositions of the targets was not measured. However the correction 
is expected to be small (less than 5\%) in view of the quoted target thicknesses and 
taking into account of the fact that $^8$B escape is partly counterbalanced by that of 
$^8$Li in these measurements which rely on $^7$Li(d,p)$^8$Li normalization.
\end{thebibliography}
\end{document}